\documentclass[sigconf]{acmart}

\usepackage{booktabs} 

\usepackage{setspace}
\usepackage{algorithm}
\usepackage{algorithmic}

\usepackage{placeins}
\usepackage{multirow}

\usepackage{subfig}

\newcommand{\fw}{FilteredWeb }
\newcommand{\efw}{\emph{\fw}}

\setcopyright{rightsretained}
\acmConference[WebSci]{10th ACM Conference on Web Science}{2018}{Amsterdam, The Netherlands} 

\begin{document}
\title{Automated Discovery of Internet Censorship by Web Crawling}


\author{Alexander Darer}
\affiliation{%
  \institution{Dept. Computer Science\\University of Oxford}
  \city{Oxford} 
  \state{UK} 
}

\author{Oliver Farnan}
\affiliation{%
  \institution{Dept. Computer Science\\University of Oxford}
  \city{Oxford} 
  \state{UK} 
}

\author{Joss Wright}
\affiliation{%
  \institution{Oxford Internet Institute\\University of Oxford}
  \city{Oxford} 
  \state{UK} 
}


\copyrightyear{2018} 
\acmYear{2018} 
\setcopyright{acmlicensed}
\acmConference[WebSci '18]{10th ACM Conference on Web Science}{May 27--30, 2018}{Amsterdam, Netherlands}
\acmBooktitle{WebSci '18: 10th ACM Conference on Web Science, May 27--30, 2018, Amsterdam, Netherlands}
\acmPrice{15.00}
\acmDOI{10.1145/3201064.3201091}
\acmISBN{978-1-4503-5563-6/18/05}

\begin{abstract}
Censorship of the Internet is widespread around the world. As access to the web becomes increasingly ubiquitous, filtering of this resource becomes more pervasive. Transparency about specific content that citizens are denied access to is atypical. To counter this, numerous techniques for maintaining URL filter lists have been proposed by various individuals and organisations that aim to empirical data on censorship for benefit of the public and wider censorship research community.

We present a new approach for discovering filtered domains in different countries. This method is fully automated and requires no human interaction. The system uses web crawling techniques to traverse between filtered sites and implements a robust method for determining if a domain is filtered. We demonstrate the effectiveness of the approach by running experiments to search for filtered content in four different censorship regimes. Our results show that we perform better than the current state of the art and have built domain filter lists an order of magnitude larger than the most widely available public lists as of Jan 2018. Further, we build a dataset mapping the interlinking nature of blocked content between domains and exhibit the tightly networked nature of censored web resources.
\end{abstract}


\maketitle

\section*{Acknowledgments}
This work was supported by EPSRC through the Centre for Doctoral Training in Cyber Security, University of Oxford
and \grantsponsor{ati}{The Alan Turing Institute}{https://www.turing.ac.uk} under the \grantsponsor{epsrc}{EPSRC}{https://www.epsrc.ac.uk} grant \grantnum{epsrc}{EP/N51012}. 

Alexander Darer \& Oliver Farnan are funded by the Center of Doctoral Training in Cyber Security.

Joss Wright is partially funded by the Alan Turing Institute as a Turing Fellow under Turing Award Number \grantnum{ati}{TU/B/000044}.

\section{Introduction}
The effort expended by censorship regimes around the world attempting to filter Internet resources they deem to be too sensitive or against the morality of their own interests is on-going. As Internet access has become more ubiquitous, the scale of deployed filtering systems is increasing. A recent study has shown that blocking the reachability of popular sites at national levels is widespread and disruptive \cite{kuhrer2015going}\cite{Pearce2017a}. Advocates for free-speech and a free Internet push for transparency and openness, while censors attempt to repress the flow of certain information within their networks. Key to this is the blocking of specific webpages and the URLs that point to them.

In response to large scale filtering of web resources, there have been numerous studies over recent years aimed at determining the type of content being blocked in different countries. Of particular interest are periods of time when blocking has occurred and the development of techniques to monitor filtered URLs and keywords \cite{crandall2007conceptdoppler}\cite{Darer2017a}\cite{fu2013assessing}\cite{knockel2011three}\cite{Sfakianakis2011a}.

We introduce an approach for discovering filtered domains in different countries at scale and reasonable cost. The system is fully automated and does not require per-country expertise or cooperation - meaning that the safety of individuals within censored regimes won't be compromised. Our method applies webcrawling techniques to find blocked content and uses a seed list of known filtered URLs to initiate the search. We make use of DNS servers within a target country as measurable devices. This allows us to monitor the filter status of individual domains and sub-domains without human intervention. The system is recursive so newly discovered filtered URLs are fed back into the search to allow on-going measurement. Results from our experiments using four different test countries have shown that our approach can be used to find filtered URLs that are not present in the original seed lists. Furthermore, we collect data about the linked nature of various filtered domains to gain further insight into how different pieces of filtered content are associated.

\subsection{Related Work}
Over the last decade there have been many approaches for detecting censorship of the Internet around the world. Of these, many are country specific and have focused on China \cite{clayton2006ignoring}\cite{fu2013assessing}\cite{king2013censorship}\cite{lowe2007great}\cite{wright2014regional}, Indonesia \cite{ooni_id}\cite{warf2011geographies}, Iran \cite{anderson2013dimming}\cite{aryan2013internet}, Pakistan \cite{aceto2016analyzing}\cite{nabi2013anatomy} and Thailand \cite{Gebhart2017a} among others.

The most widely adopted and current URL filter lists are maintained by the \emph{CitizenLab} \cite{czlab_lists}. They are constructed using local knowledge and reports of filtering in different countries and collate data from different sources such as \emph{OONI} \cite{filasto2012ooni}.

Developing new techniques for discovering filtered URLs is a challenging problem. Yet, this is a rich research field with numerous techniques published over recent years \cite{Aceto2015b}. The use of DNS as a means of testing censorship of web content is not new, but can be advantageous due to its scalability and remote nature \cite{wander2014measurement}. These attributes make DNS a common tool for other censorship monitoring architectures such as \emph{UBICA} \cite{Aceto2015a}, \efw \cite{Darer2017a} and \emph{CensMon} \cite{Sfakianakis2011a}.

Building in-depth and accurate URL filter lists is an important aspect for censorship research. These collections are in widespread use among the research community for various different measurements and tests for internet reachability, web content blocking and circumvention techniques \cite{Pearce2017b}\cite{Weinberg2017a}. Furthermore, the subsequent and on-going maintenance of these lists provides opportunities for insight into the condition of internet filtering around the world. The data collected by the aforementioned monitoring architectures is vital if we are to construct a model of censorship as it develops.

\subsection{Contributions}
This paper introduces a new approach for discovering filtered domains within target censorship regimes. We have created an implementation of the technique and, through experimentation, shown it to be an effective tool for building URL filter lists. Furthermore, our results reveal that the approach has found significantly more filtered URLs for the test countries than were currently available in the largest public filter lists. Our formalised contributions are:
\begin{itemize}
\item A new approach for discovering previously unknown filtered domains
\item Experimental analysis of the technique through measurement of filtering activity within four know censorship regimes
\item Category breakdown of the types of content being blocked within these regimes
\item Analysis of forward filtered links and filtered backlinks of webpages on filtered domains
\end{itemize}

A substantial research output from this body of work is a test list containing a large number of currently filtered domains within China, Indonesia, Iran and Turkey that have previously been unpublished. We aim to make this list available as soon as possible to the wider censorship research community.

\section{Traversal of Filtered Webpages}
Traversal between webpages using embedded hyperlinks is the most widely used method for content discovery on the Web. Our aim is to exploit the connections between different filtered webpages to efficiently crawl sites in search for more blocked content. An important assumption for this approach is that different filtered webpages do indeed link to others. We demonstrate this through experimental analysis of four different countries that are known to filter websites via DNS manipulation. The method we describe is not dissimilar to conventional web crawling techniques widely used by large search engines. Given this, we aim to build a new dataset that contains information pertaining to backlinks\footnote{Backlinks are hyperlinks that point to a certain page from other pages.} of filtered webpages.

\begin{figure}[!t]
\centering
\includegraphics[width=0.48\textwidth]{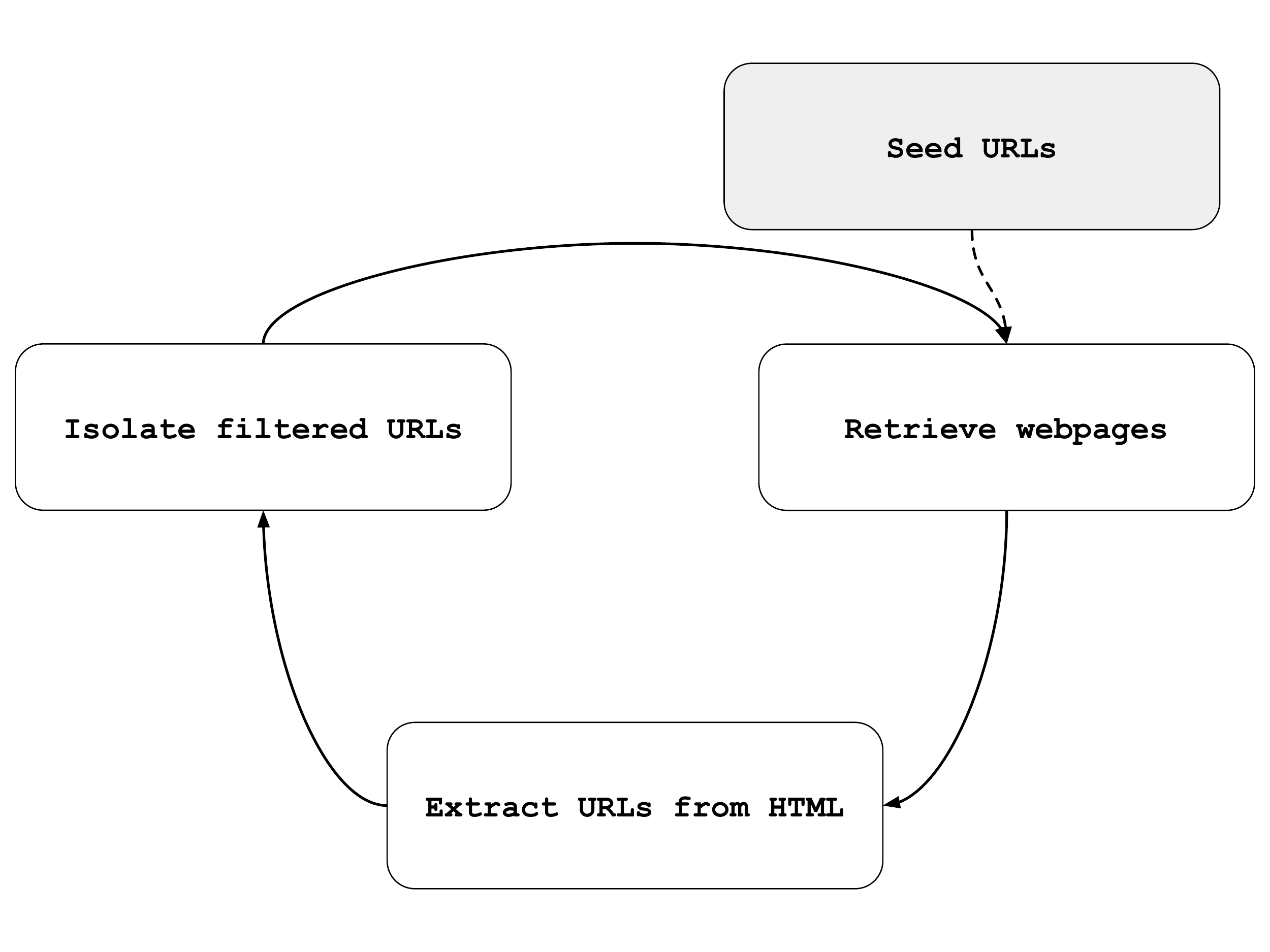}
\caption{High-level overview of filtered webpage traversal}
\label{fig:snowflow}
\end{figure}

This technique works on a simple premise - filtered webpages contain links to other filtered webpages. We begin the discovery by seeding the system with a number of known filtered URLs with the presupposition that these will contain hyperlinks to further blocked content. A high-level overview of the technique is shown in Figure~\ref{fig:snowflow} and works as follows:

\begin{enumerate}
\item Start with a list of known filtered URLs for country $c$
\item Retrieve webpages for all known filtered URLs in our list
\item Extract any URLs from the downloaded webpages
\item Isolate the URLs that are filtered in country $c$ from the extracted URLs
\item Add the newly identified filtered URLs to the list, then goto step 2
\end{enumerate}

A number of hyperlinks in any webpage will point to resources that do not provide utility for our discovery. We ignore any URLs that point to static HTML assets - such as javascript, css or image files and also remove any self-referencing URLs - hyperlinks to the same domain for the webpage. We aim to reduce the possibility of having the crawler becoming stuck in cliques such as affiliate or adult site networks this way. For purposes of analysis of the approach we visit each unique URL only once.

\subsection{Methods for Checking Webpage Availability}
A key part of this system is the capability to determine if a domain is filtered in a certain country or not. This is important as crawling unnecessarily large portions of the Web will make the discovery inefficient. Furthermore, an inaccurate checking procedure could produce large numbers of false positives - detracting from the usefulness of the results. We must also be prudent in regards to ethical issues when taking internet measurements using such a system. Since, we will be probing for blocked content, we want to ensure that the safety of individuals is not compromised as described by \cite{dittrich2011menlo}.

Given these requirements, we employ a checking system that uses DNS infrastructure to determine if domains are being filtered within a given country. The exploitation of DNS as a means for blocking access to certain web resources is in widespread use around the globe \cite{farnan2016poisoning}\cite{kuhrer2015going}\cite{lowe2007great}\cite{Pearce2017b}; and because resolvers operated by large Internet Service Providers (ISPs) are often open, we can use these as a basis for our measurements.

To determine the filter status of a domain, we require a globally non-censored DNS server as a control and a measurement DNS server located within a target country. Using these, we run through a process to comprehensively check if a domain is filtered by the measurement server. We examine responses to DNS queries made for the test domain to determine if the server is poisoned \cite{farnan2016poisoning} or acting rogue. This procedure is described in Algorithm~\ref{alg:filtercheck}. 

The following six checks are used to ascertain the filter status of domains:
\begin{enumerate}
\item A DNS query is intercepted in the target country when sent to a non-existent DNS server
\item The measurement server times out but the control server does not
\item The measurement server responds with a private IP address but the control does not
\item The measurement server resolves an IP that times out on an HTTP GET request, whereas the resolved IP from the control server does not
\item The control server resolves an IP that times out on an HTTP GET request, whereas the resolved IP from the measurement server does not
\item The content length of the webpages from each resolved IP differs by more than a defined percentage amount
\end{enumerate}

If the results from any of the above are found to be positive, we consider the domain to be filtered by the measurement DNS server. 

\begin{algorithm}
\caption{Pseudocode for Domain Filtering Check}
\label{alg:filtercheck}
\begin{algorithmic}[1]
\item[$MAXDIFF \Leftarrow p$]\COMMENT{content length \% difference that indicates filtered domain}

\item[$dom$]\COMMENT{domain to check}
\item[$mDNS$]\COMMENT{measurement DNS server in target country}
\item[$mDNSFake$]\COMMENT{fake DNS server in target country}
\item[$cDNS$]\COMMENT{control DNS server}
\item[]

\item[]\COMMENT{the following variables are $NULL$ on timeout}
\STATE $mFakeIP \Leftarrow resolveDNS(dom, mDNSFake)$
\STATE $mIP \Leftarrow resolveDNS(dom, mDNS)$
\STATE $cIP \Leftarrow resolveDNS(dom, cDNS)$
\STATE $mIPContent \Leftarrow httpGETRequest(mIP)$
\STATE $cIPContent \Leftarrow httpGETRequest(cIP)$
\item[]

\IF {$mFakeIP \neq NULL$}
  \RETURN{$TRUE$} \COMMENT{DNS query was intercepted in target country}
\ENDIF
\item[]

\IF{$mIP == NULL$ \AND $cIP \neq NULL$}
  \RETURN{$TRUE$} \COMMENT{$mDNS$ is rogue server}
\ENDIF
\item[]

\IF{$mIP \neq cIP$}
  \IF{$isPrivateIP(mIP) == TRUE$ \AND $isPrivateIP(cIP) == FALSE$}
    \RETURN{$TRUE$} \COMMENT{$mIP$ is private address}
  \ENDIF
  \IF{$mIPContent == NULL$ \AND $cIPContent \neq NULL$}
    \RETURN{$TRUE$} \COMMENT{$mDNS$ is rogue server}
  \ENDIF
  \IF{$mIPContent \neq NULL$ \AND $cIPContent == NULL$}
    \RETURN{$TRUE$} \COMMENT{$mDNS$ is rogue server}
  \ENDIF
  \IF{$length(mIPContent)/length(cIPContent) > MAXDIFF$}
    \RETURN{$TRUE$} \COMMENT{$mIP$ points to incorrect content}
  \ENDIF
\ENDIF
\RETURN{$FALSE$} \COMMENT{$dom$ not filtered}

\end{algorithmic}
\end{algorithm}

This procedure has a number of useful features. Firstly, it does not require cooperation of any person or individual within a censored country, it solely makes measurements on infrastructure. Second, it is an efficient mechanism that is scalable and yields fast results. Third, we can perform measurements from outside target countries giving us the capability to analyse filtering within a wide array of censorship regimes.

The limitation with this approach is that we lose fine grain information about individual URLs that may be blocked. This is due to the sole use of DNS servers as a checking mechanism - since one can only query about entire domains or sub-domains. Further, we require that measurement DNS servers respond to queries from remote countries in the same manner they do for queries made domestically. Yet, as we need to make a trade-off between effectiveness, efficiency and ethical considerations, this method is sufficient for use to generate good results to show the usefulness of this technique for discovering filtered domains as a whole.

\subsection{Ethical Considerations}
Implications of censorship measurements open us up to a number of ethical issues that we must give thought to. First and foremost, it is imperative that we do not cause harm to any persons or organisations that are unaware of our actions and motivation. This can be casual in many ways, not least because testing of internet filtering often requires sending network traffic to and from censorship regimes \cite{Crandall2015a}. Certain studies within the field have required the use of aware volunteers who are located within countries of interest. While these individuals are generally knowledgeable of the motivation of the study and potential ramifications of their actions if implicated, this is not something we as researches should take lightly. In many cases, it is simply not appropriate to use human participants for this type of work. Furthermore, there are a number of legal issues with measurements of censorship based on the techniques used - especially if inference is made using direct observations within a target country \cite{Wright2011a}.

We must also consider the use and deployment of these kinds of discovery techniques by antagonists. Since we aim to build a system that can automatically find alternative content that is blocked based on \emph{known} blocked content, such a framework could be utilised in an adverse way to filter further web resources. Unfortunately, we cannot guarantee that this use-case will never occur given the fact that censors generally do not publish technological details about their infrastructure and systems.

These concerns should not however reduce our willingness to practice this kind of research. If considerable effort is made to ensure our measurements will not affect individuals, we are able to provide empirical data concerning censorship around the world. This can give us as researchers a substantial insight into complex socio-political issues that are of benefit to the community and are of wider public interest given the state fragile of international relations. Moreover, our proposed approach does not pose a risk to individuals or rely human volunteers and vulnerable subjects. We take measurements directly from infrastructure in a manner that the services were originally designed for.

\section{Experimental Analysis}
We conduct experiments on four different countries with an aim to build domain filter lists that are longer and more in-depth than are currently available. This was achieved using an implementation of the approach written in \emph{Python} with the following parameters:
\begin{itemize}
\item Control DNS server: 8.8.8.8
\item $MAXDIFF$ {\scriptsize(content-length difference that indicates filtering)}: 50\%
\item Filter check timeout: 10 seconds
\item Maximum recursion depth\footnote{This is the maximum depth of recursion from the seed URLs}: 100
\item Seed URLs obtained from the CitizenLab filter lists \cite{czlab_lists}
\end{itemize}

The $MAXDIFF$ value is used based on a study that found the content-length of censorship block pages are 95\% likely to differ by more than 50\% compared to the genuine page \cite{aceto2014monitoring}. Further, we ensure that the system does not follow links that self-reference the parent site - this is to say we attempt to stop looping behaviour with pages that link to others on the same domain. Also, we never revisit a URL that has previously been seen - it will be counted in the statistics we gather, but not checked again.

The target countries tested were: China, Indonesia, Iran and Turkey. Each experiment ran for seven days, or until no more filtered domains were found. The DNS servers used for each test country are shown in Table~\ref{table:dnsservers}. The real DNS servers were selected from large ISPs in the target countries and the fake from the pool of unallocated IP addresses also owned by the same ISPs. We do this because as mentioned previously, we aim to take measurements on mass infrastructure within the target countries rather than smaller organisations or individuals.

\begin{table*}[!t]
  \renewcommand{\arraystretch}{1.3}
  \caption{DNS servers used for experiments}
  \label{table:dnsservers}
  \centering
  \begin{tabular}{|l||c c| c |}
    \hline
    & \textbf{Real Servers} & \textbf{Fake Servers} & \textbf{ISP} \\ \hline \hline
    \multirow{2}{*}{\textbf{China}} & 202.46.32.29 & 220.181.57.217 & \multirow{2}{*}{Shenzhen Sunrise Technology Co. Ltd.} \\
    & 180.76.76.76 & 223.96.100.100 & \\ \hline
    \multirow{2}{*}{\textbf{Indonesia}} & 202.134.0.155 & 202.134.2.10 & \multirow{2}{*}{PT Telkom Divisi Multimedia} \\
    & 202.134.1.10 & 180.131.144.44 & \\ \hline
    \multirow{2}{*}{\textbf{Iran}} & 94.183.43.170 & 94.183.92.90 & \multirow{2}{*}{Aria Shatel Company Ltd} \\
    & 2.179.167.100 & 5.161.128.10 & \\ \hline
    \multirow{2}{*}{\textbf{Turkey}} & 195.175.39.39 & 195.175.30.39 & \multirow{2}{*}{Turk Telekomunikasyon Anonim Sirketi} \\
    & 195.175.39.40 & 195.175.30.100 & \\ \hline
  \end{tabular}
\end{table*}

\subsection{Results}
Table~\ref{table:results} depicts the number of unique URLs extracted over the course of each experiment and how many of those were filtered in the given country. We also perform a count on the number of unique filtered domains within the list of filtered URLs. As a measure for the breadth of each run, the Alexa Top 1000 domains were removed so we can analyise how deep the system is able to penetrate to lesser known sites with lower numbers of visitors and backlinks.

In total, we extracted over 80 million URLs from filtered web pages, of which 5.7 million were themselves from a filtered domain. The number of blocked domains identified for Turkey and Indonesia are an order of magnitude larger than those found for China and Iran. This is due to the widespread censorship of adult related sites within these particular censorship regimes. Turkey passed a law in 2007 prompting the explicit blocking of over 80,000 sites, of which many contained adult content \cite{akgul2015internet}, and Indonesia, a similar ban in 2010 \cite{indi_blocks_porn} \& 2017 \cite{indi_blocks}.

We perform a comparison with the most widely available public URL filter lists, maintained by the CitizenLab. To ensure a fair comparison, we run these lists through our filtering check and report those numbers. The figures are shown in Table~\ref{table:compare}. From this we can show that we have performed efficiently and identified more filtered domains than were present in the original seed lists. To gain further insight into the types of content filtered in Turkey and Indonesia, we remove the adult domains to create separate counts for better comparison.

Our results demonstrate that this approach is effective at finding previously unknown filtered domains. A major advantage of this technique is that \emph{only} URLs from filtered domains are visited, meaning that we can achieve efficient webcrawling.

\begin{table*}[t]
  \renewcommand{\arraystretch}{1.3}
  \caption{Results from experimental analysis}
  \label{table:results}
  \centering
  \begin{tabular}{|l||c c || c c|}
    \hline
    & \multirow{2}{*}{\textbf{Extracted URLs}} & \multirow{2}{*}{\textbf{Filtered URLs}} & \multirow{2}{*}{\textbf{Filtered Domains}} & \multirow{2}{*}{\textbf{Filtered Domains}} \\ 
    & {\footnotesize \emph{(HTML assets and self-linking URLs removed)}} & & & {\footnotesize \emph{(Alexa Top 1000 removed)}} \\ \hline \hline
    \textbf{China} & 33,082,217 & 2,098,264 & 1576 & 1454 \\ \hline
    \textbf{Indonesia} & 12,580,357 & 835,395 & 47,143 & 47,065 \\ \hline
    \textbf{Iran} & 15,381,873 & 1,868,852 & 651 & 576 \\ \hline
    \textbf{Turkey} & 19,250,931 & 913,213 & 39,725 & 39,614 \\ \hline \hline
    \textit{Totals:} & 80,295,378 & 5,715,724 & 89,095 & 88,709 \\ \hline 
  \end{tabular}
\end{table*}

\begin{table}[t]
  \renewcommand{\arraystretch}{1.3}
  \caption{Comparison of results to CitizenLab filter lists\\ \emph{\small CitizenLab figures accurate as of 1st Sept 2017}}
  \label{table:compare}
  \centering
  \begin{tabular}{|l||c c|}
    \hline
    & \multicolumn{2}{|c|}{\multirow{2}{*}{\textbf{Filtered Domains}}} \\ 
    & \multicolumn{2}{|c|}{\footnotesize \emph{(Alexa Top 1000 removed)}} \\ \hline \hline
    & \emph{CitizenLab} & Darer et al. \\ \hline
    \textbf{China} & 127 & 1454 \\ \hline
    \textbf{Indonesia {\footnotesize\emph{(Adult domains removed)}}} & 124 & 1280 \\ \hline
    \textbf{Iran} & 351 & 576 \\ \hline
    \textbf{Turkey {\footnotesize\emph{(Adult domains removed)}}} & 131 & 513 \\ \hline
  \end{tabular}
\end{table}

\section{Further Analysis}
The experiments we performed have yielded an interesting dataset that lends itself to further investigation. We are able to track the paths that lead to filtered content by analysing routes taken by the crawler. This gives us a useful base for examining how deeply connected collections of blocked sites are. Further, we can identify the backlinks of filtered pages and the outbound (forward) links to other filtered sites, we can discover networks of filtered sites.

We find that the results found in Turkey and Indonesia contain large numbers of adult sites - which as explained, are known to be banned. This observed behaviour of our tool may be due to the way that adult websites and businesses associate their domains together with the use of vast networks of traffic brokers, domain redirectors and link collections \cite{wondracek2010internet}. Based on this networking effect the webcrawler may traverse content within this subject matter given the tightly linking nature of the sites - site A references site B and site B references site A, etc. However, this is important behaviour for this approach because different pages within each site may contain distinct filtered URLs. The limitation is that the crawler may get stuck in a loop within a closed network. Even so, our results contain over 1292 filtered non-adult domains for Indonesia and 528 filtered non-adult domains for Turkey.

The results for China and Iran show significant improvement over the original seed lists of filtered domains, with our number for China over 10 times greater than the input to the system and Iran over 60\% higher.


\subsection{Top-Level-Domain Enumeration}
We perform an enumeration of all publicly available top-level-domains (TLDs) that can be attributed to different domains - and therefore different DNS records. We use the Public Suffix List maintained by the Mozilla Foundation \cite{pubsuf}. This list of TLDs contains all known public suffixes, common examples such as \emph{.com} and \emph{.org}, and less well-known instances such as \emph{pvt.k12.ma.us}. For each filtered domain discovered in a target country, we remove the TLD and check the domain, along with any subdomains, with all suffixes in the list for filtering in that country. For Indonesia and Turkey, we run the test on the non-adult domains only for better comparison. Results of the enumeration are shown in Table~\ref{table:tldenum}.

\begin{table}[t]
  \renewcommand{\arraystretch}{1.3}
  \caption{Filtered domain counts after TLD enumeration}
  \label{table:tldenum}
  \centering
  \begin{tabular}{|l||c c|}
    \hline
    & \textbf{Filtered Domains} & \textbf{Of which, hosts exist} \\ \hline \hline
    \textbf{China} & 97,167 & 5408 \\ \hline
    \textbf{Indonesia} & 1479 & 1543\\ \hline
    \textbf{Iran} & 5970 & 4527 \\ \hline
    \textbf{Turkey} & 789 & 584 \\ \hline
  \end{tabular}
\end{table}

Having completed this process, we find a large number of alternative TLDs for the filtered domains discovered through the traversal are also themselves filtered. During this process, we find that many of the enumerated domains found to be blocked by DNS in the target countries do not have records associated with them held by the control server. In particular, 94\% of the enumerated domains found to be filtered in China received NXDOMAIN responses from the control which could therefore not resolve them. A reason for this could be that censored websites may be "retired" or move onto new domains and hosting infrastructure to evade the block. While this is a case for completely removing them from the set of results presented here, they are still explicitly filtered within the country - showing that the authorities continue to block access to them. This could be due to the stance of the censorship regime or the fact that once a site is filtered, the process for removing them from blacklists is less than trivial.

\subsection{Categories of Filtered Domains}
\begin{figure*}[!t]
\centering
\includegraphics[width=1\textwidth]{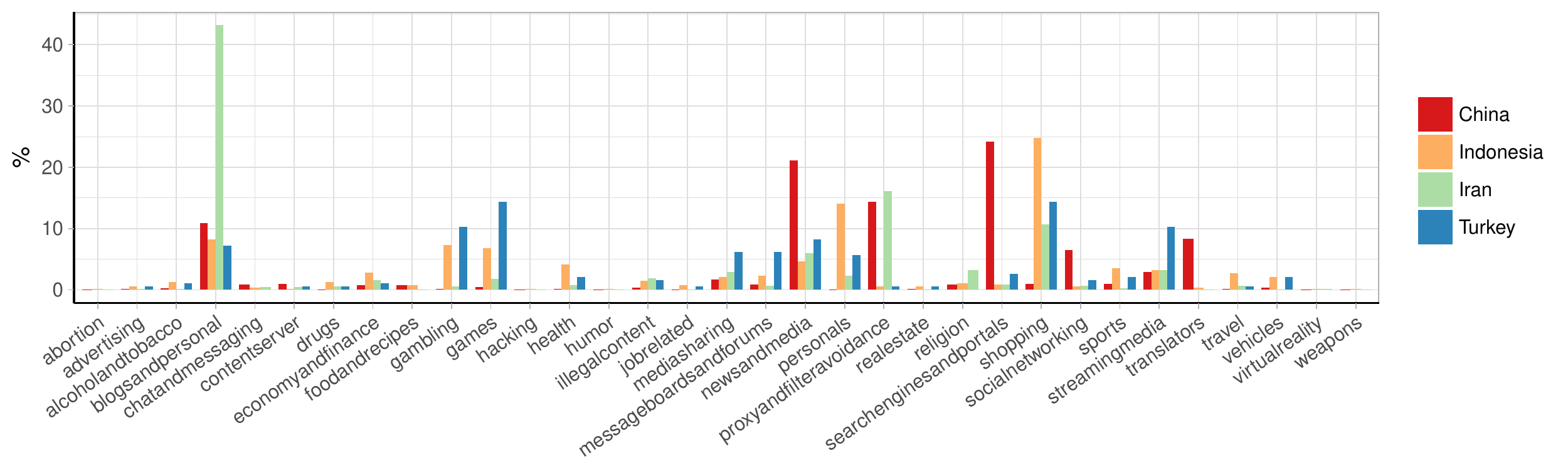}
\caption{Category breakdown of filtered domains for each target country}
\label{fig:cat_country}
\end{figure*}

To gain insight into the types of content being blocked, we run a category analysis on our list of filtered domains using the WebShrinker Categories API \cite{webshrinker}. This returns a list of categories attributed to each domain and allows us to isolate from a high level different genres of websites that are being blocked in each target country. Figure~\ref{fig:cat_country} shows the breakdown of categories for each country. From this we can see that certain types of site are overwhelmingly being blocked over others. Of particular interest is the filtering of news and media, search engines and translators by China, personals and shopping by Indonesia and games and streaming media by Turkey. We also note that the proportion of proxy and filter avoidance sites blocked by China and Iran to be comparatively high too. This is in line with recent statements from the Chinese government concerning mandatory blocking of VPNs by network providers in the country \cite{bloomvpnchina} and a similar circumstance around the Iranian presidential election in 2013 \cite{reutersiranvpn}.

\begin{figure*}[!t]
\centering
\includegraphics[width=1\textwidth]{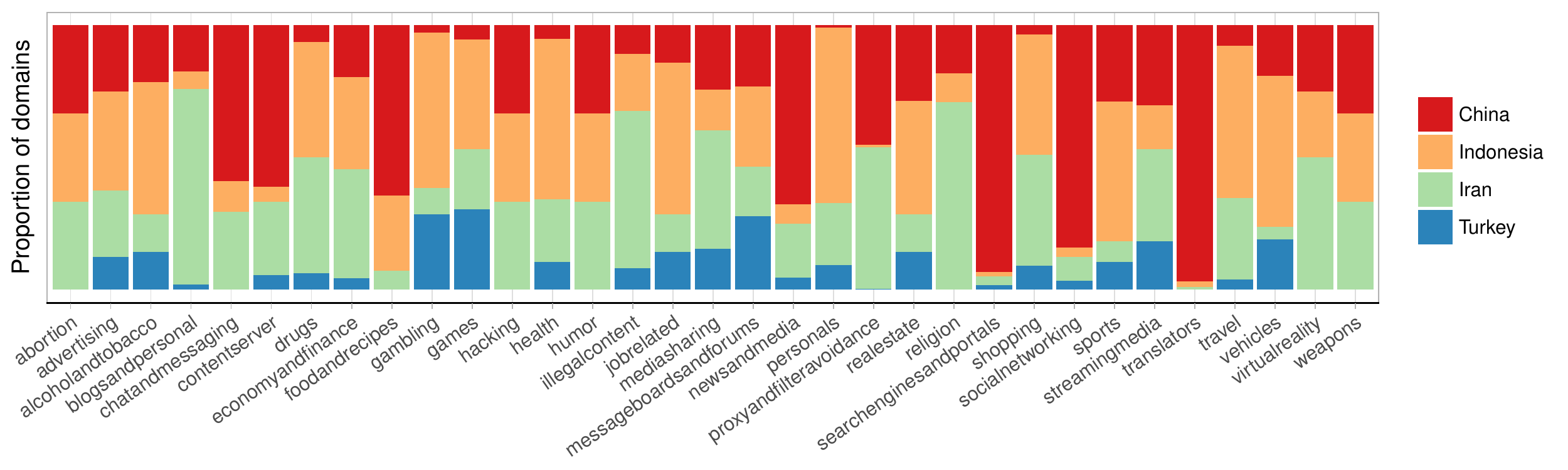}
\caption{Category comparison of filtered domains between target countries}
\label{fig:cat_prop}
\end{figure*}

Figure~\ref{fig:cat_prop} shows a comparison of categories of the filtered domains between the four test countries. This is the proportion of filtered domains per category per country. From this we can infer the different types of content that are under attention by the different regimes. For example, filtering of content within the topic of weapons is even between China, Indonesia and Iran, however censorship of religious sites is more prevalent in Iran.

\subsection{Geographical location of blocked hosts}
\begin{figure*}[!t]
\centering
\includegraphics[width=1\textwidth]{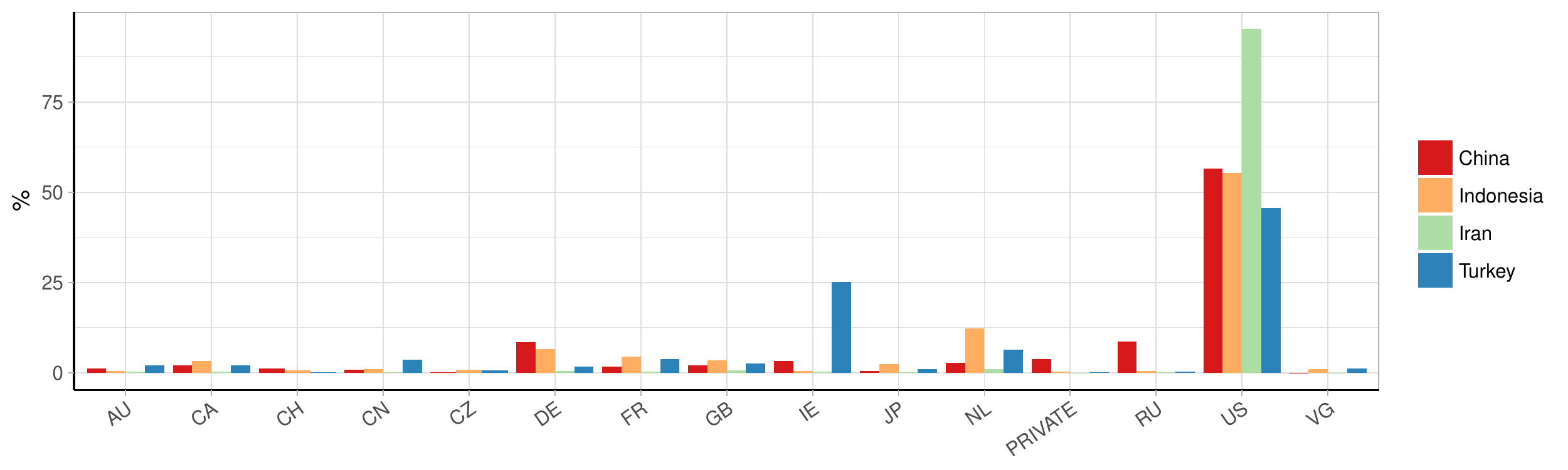}
\caption{Location breakdown of hosts serving filtered domains for each target country}
\label{fig:loc_norm}
\end{figure*}

In addition to inferring the types of content being blocked, we identify the locations of the servers hosting filtered domains in each test country. This is achieved by making a DNS query for each domain to the control server and using MaxMind GeoIP2 country database \cite{maxmind} to locate the resulting IP addresses by country of origin. The breakdown of the origin of hosts of filtered domains to test country is shown in Figure~\ref{fig:loc_norm}.

Unsurprisingly, we find that the largest number of servers are hosted within the United States. This is expected due to the way many content-delivery-networks maintain peers in North America and the fact that over 50\% of all Internet hosts are located on this continent \cite{inethosts}.

During the course of this investigation we observe that a disproportionate percentage of blocked domains for Turkey were hosts in the Republic of Ireland. On further analysis of the domains and IP address records we find that the country appears to block any subdomain of \emph{evennode.com} which is a hosting provider for NodeJS and Python web applications. The IP addresses of the blocked domains are owned by Amazon Technologies Inc. as part of their datacentres supporting Amazon Web Services. Further examination of this peculiarity was not performed, but it opens the questions as to whether certain censorship regimes will filter entire blocks of IP addresses and domains based on their hosted locations.

Other cases of interest are the irregular blocking of Dutch sites by Indonesia and Russian sites by China.

\subsection{Backlink Analysis}
For a more in-depth look into the networking effect between blocked websites, we find the number of filtered backlinks to and filtered forward links from each blocked webpage\footnote{Note that the backlinks and forward links are also themselves filtered in the given target country}. This allows us to see how deeply integrated each censored site is within the network of filtered content. We can look at the number of sites referencing \emph{a given} blocked domain and also which filtered sites reference the most \emph{other} blocked domains. 

\begin{figure*}[!t]
\centering
\subfloat[]{\includegraphics[width=0.45\textwidth]{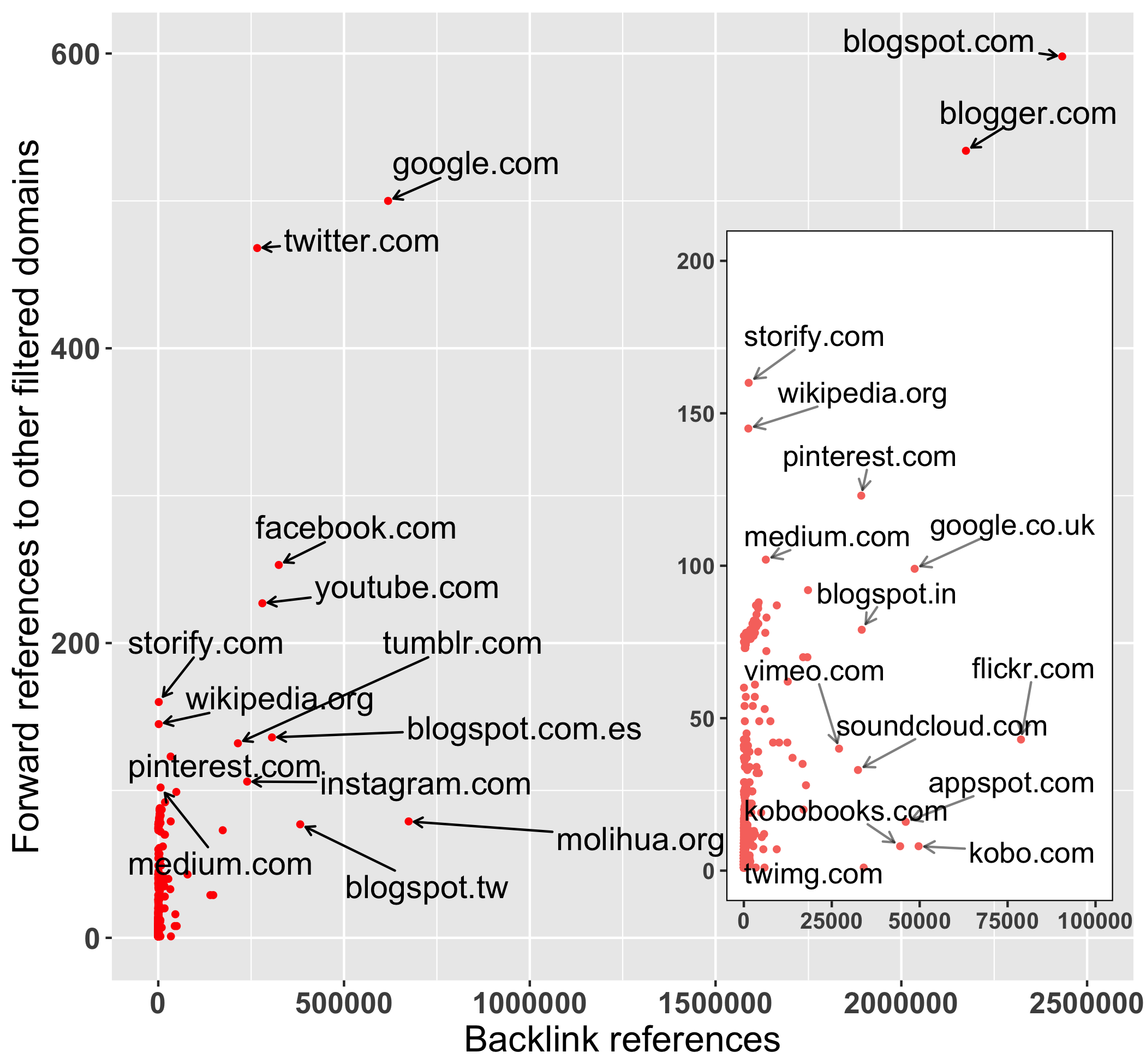}
\label{fig:backlink_china}}
\hfil
\subfloat[Top 1000 removed]{\includegraphics[width=0.45\textwidth]{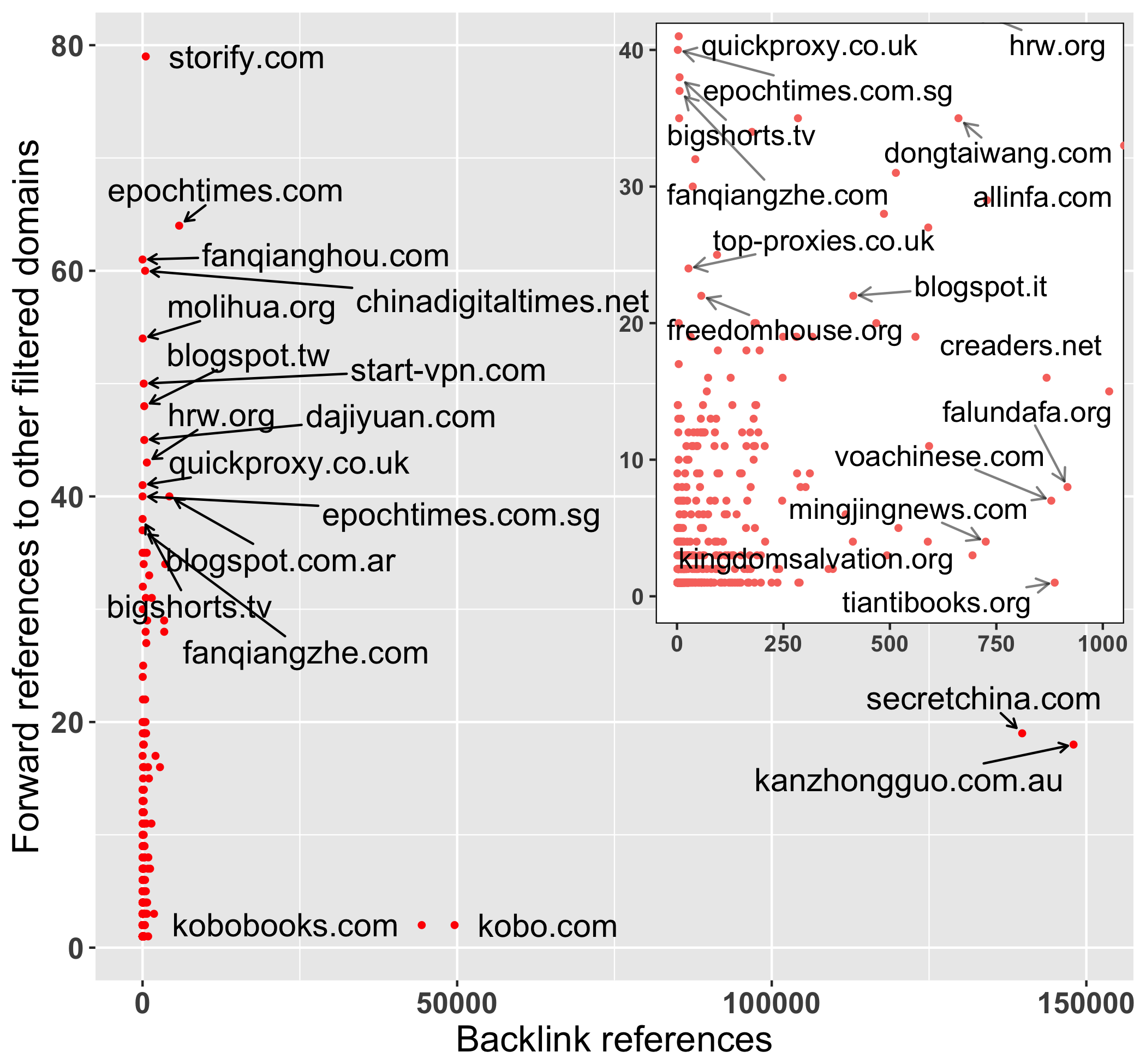}
\label{fig:backlink_china_top_1000}}
\caption{Backlinks of discovered filtered domains - China}
\label{fig:backlinks_cn}
\end{figure*}

\begin{figure*}[!t]
\centering
\subfloat[]{\includegraphics[width=0.45\textwidth]{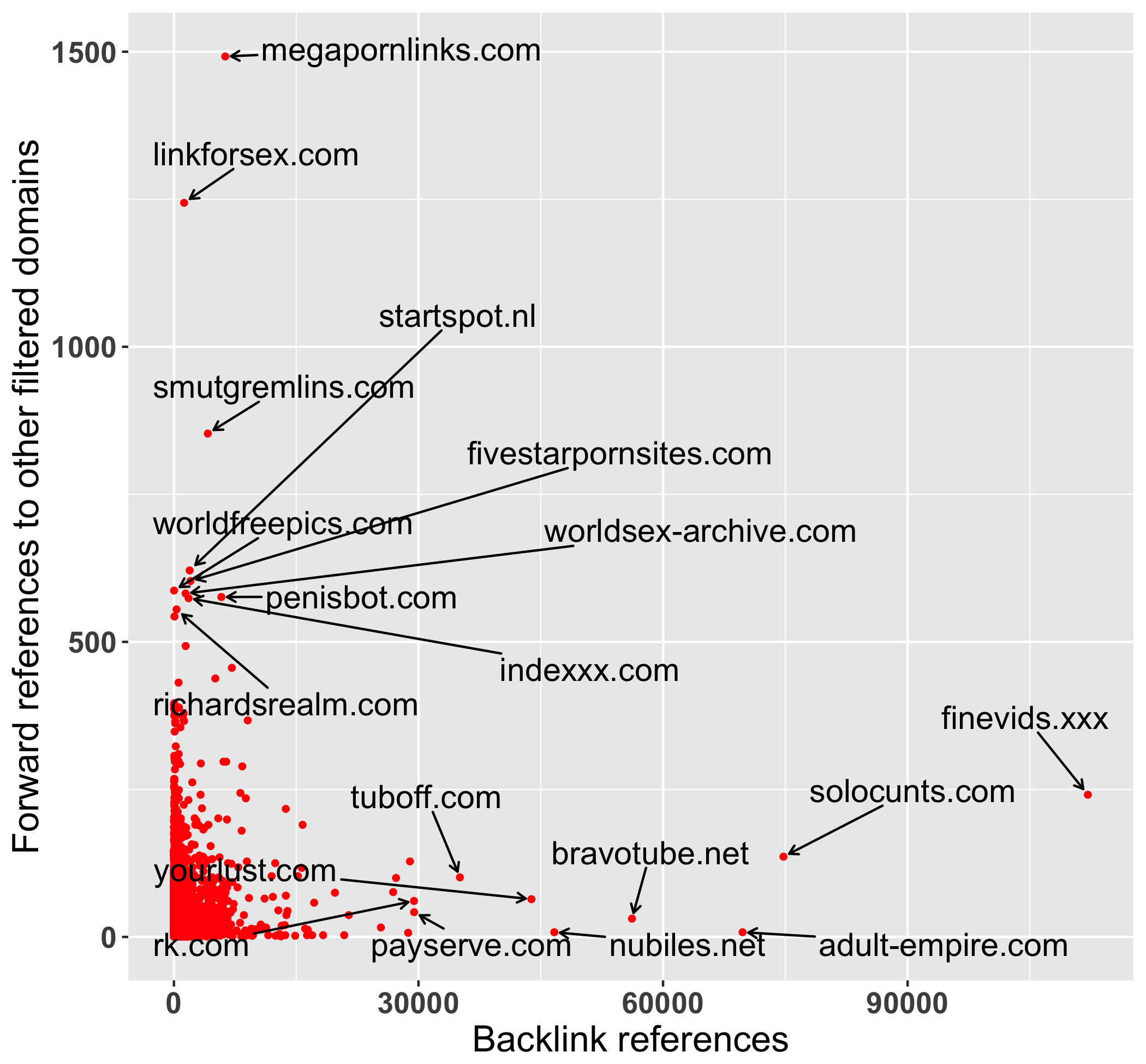}
\label{fig:backlink_indonesia}}
\hfil
\subfloat[Adult sites removed]{\includegraphics[width=0.45\textwidth]{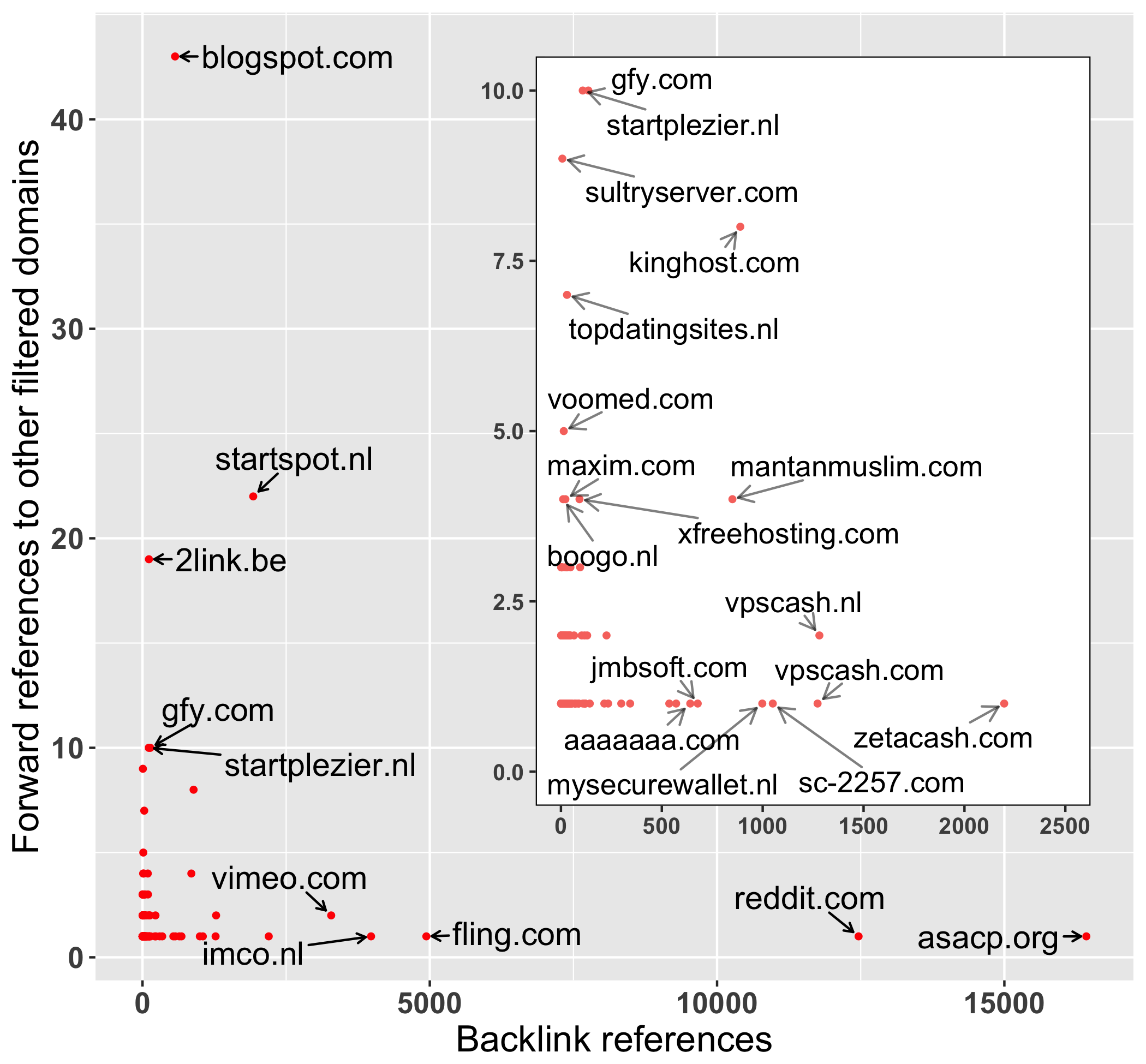}
\label{fig:backlink_indonesia_porn_removed}}
\caption{Backlinks of discovered filtered domains - Indonesia}
\label{fig:backlinks_id}
\end{figure*}

\begin{figure*}[!t]
\centering
\subfloat[]{\includegraphics[width=0.45\textwidth]{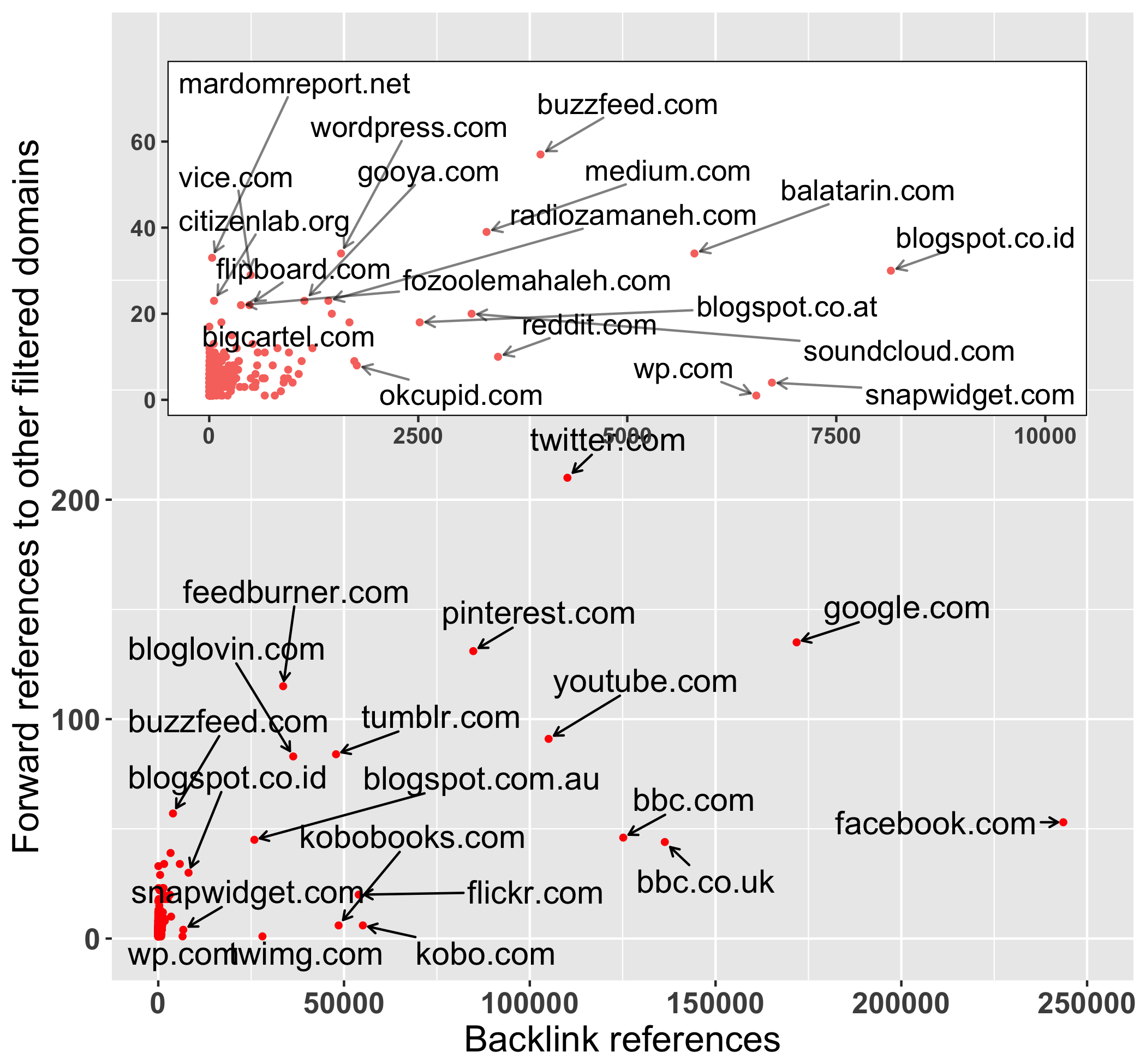}
\label{fig:backlink_iran}}
\hfil
\subfloat[Top 1000 removed]{\includegraphics[width=0.45\textwidth]{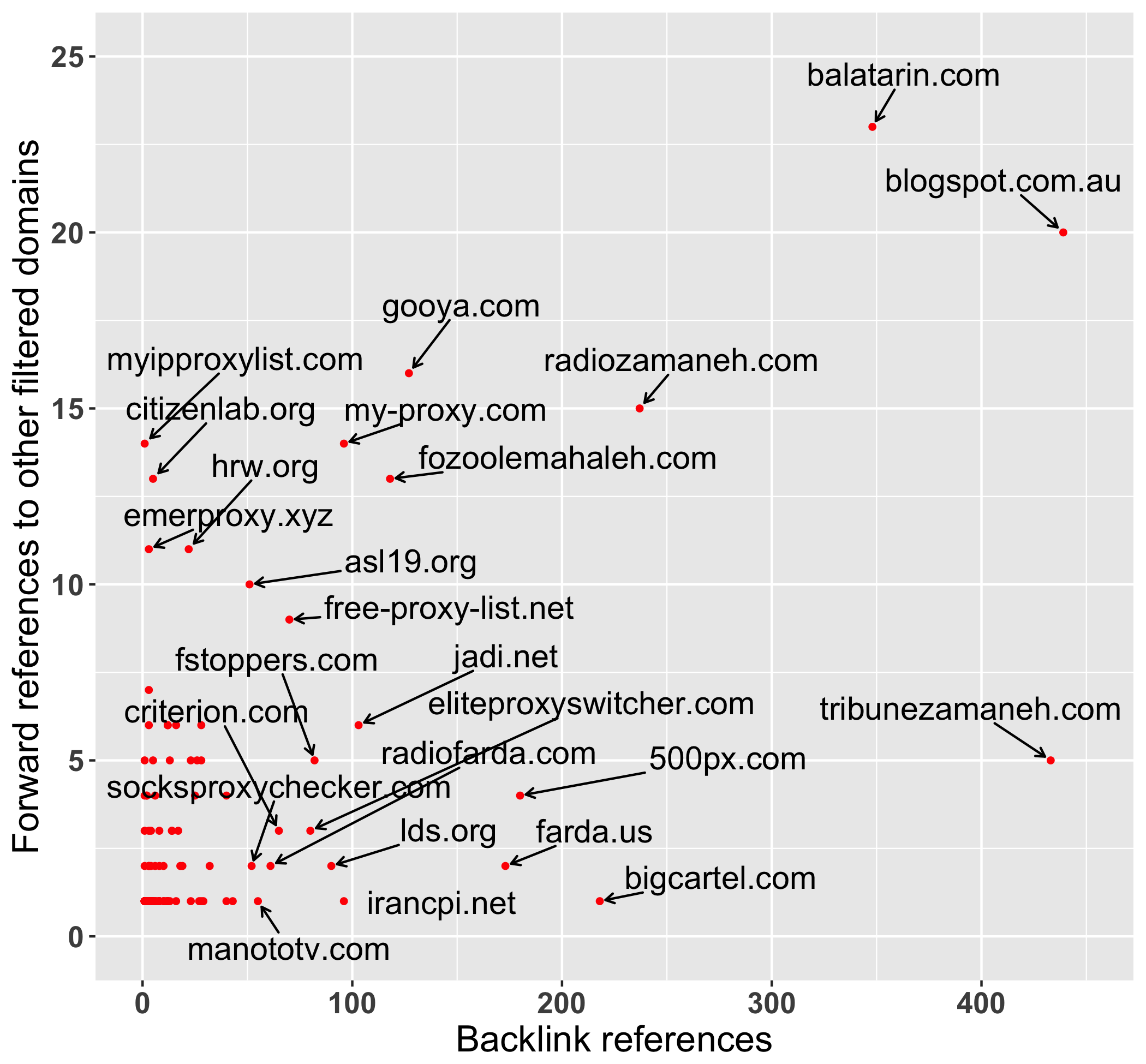}
\label{fig:backlink_iran_top_1000}}
\caption{Backlinks of discovered filtered domains - Iran}
\label{fig:backlinks_ir}
\end{figure*}

\begin{figure*}[!t]
\centering
\subfloat[]{\includegraphics[width=0.45\textwidth]{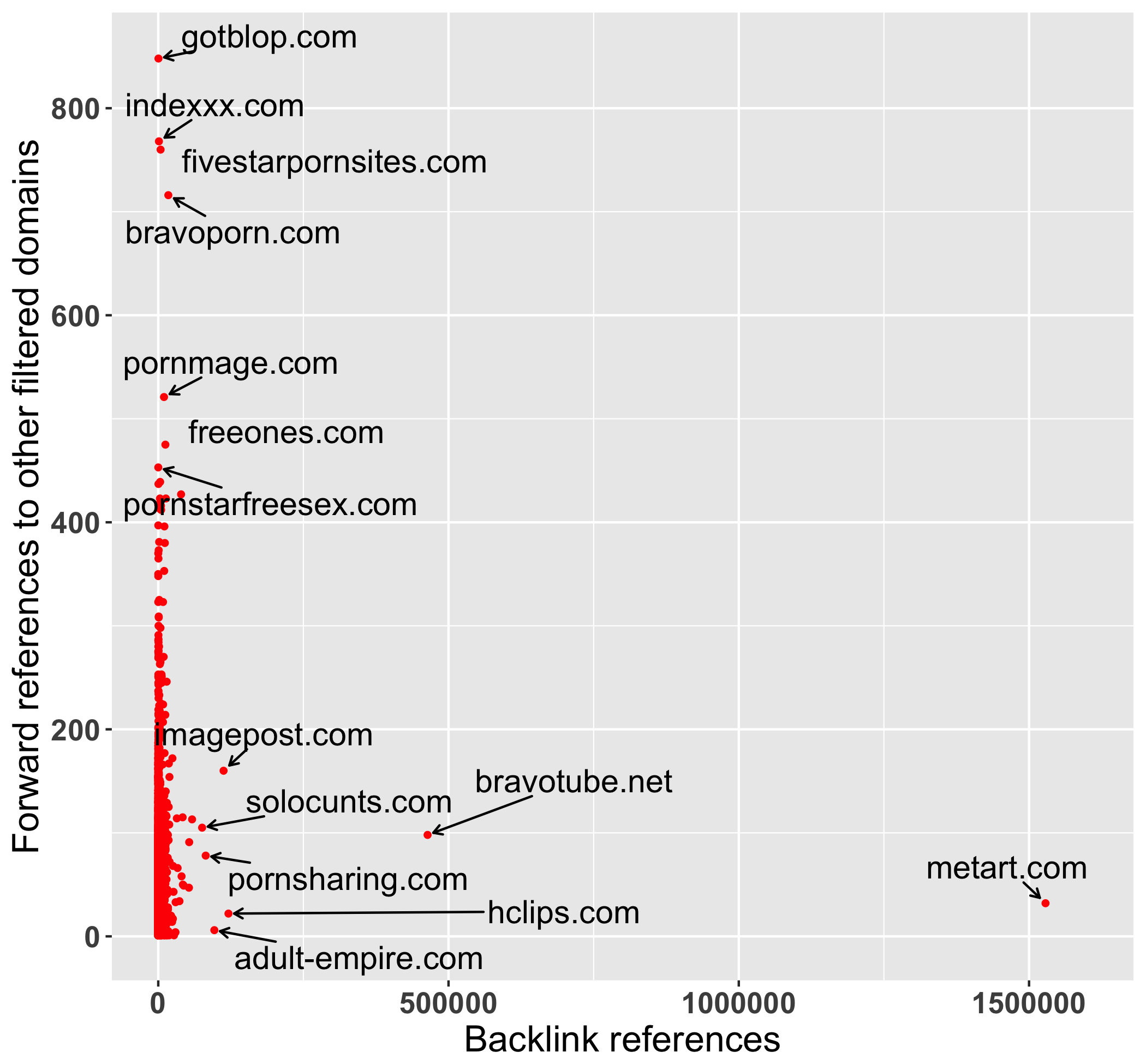}
\label{fig:backlink_turkey}}
\hfil
\subfloat[Adult sites removed]{\includegraphics[width=0.45\textwidth]{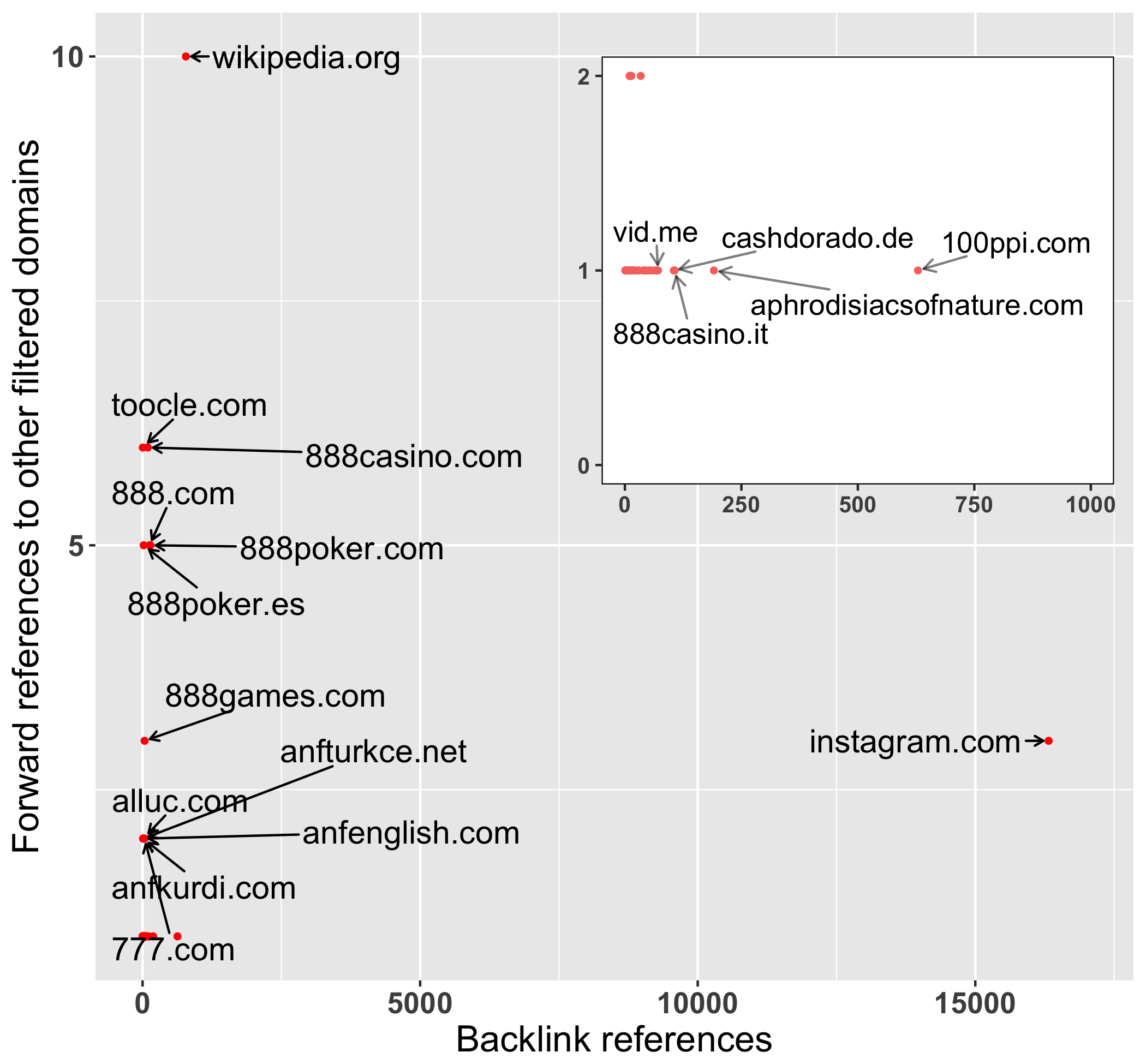}
\label{fig:backlink_turkey_porn_removed}}

\caption{Backlinks of discovered filtered domains - Turkey}
\label{fig:backlinks_tr}
\end{figure*}

To calculate these, we log every backlink we find to a filtered domain along with the filtered domains found to be linked \emph{from} each filtered domain (forward filtered links). This results in a large graph of interconnected nodes (where each node is a filtered domain) and edges representing hyperlinks between them. From this, we can gain an insight into which domains are highly referenced within the network and which domains contain the most references to other filtered domains. Figures~\ref{fig:backlinks_cn}, \ref{fig:backlinks_id}, \ref{fig:backlinks_ir} and \ref{fig:backlinks_tr} show the backlinks of filtered domains for each target country.

Notable observations in Figures~\ref{fig:backlink_indonesia} and \ref{fig:backlink_turkey} are that the top sites that link to other filtered domains appear to be adult link collections which supports the findings in \cite{wondracek2010internet}. We can also see in Figures~\ref{fig:backlink_china_top_1000} and \ref{fig:backlink_iran_top_1000} that many of the linkers to filtered content are freedom of expression and independent news sites, both of which often contain political criticism.

\subsection{Limitations}
As mentioned previously, the filter check is limited by the sole use of DNS. While this reduces cause for ethical concerns, it does mean that content filtered by other means - such as IP filtering, keyword filtering or Deep Packet Inspection - will not be marked as blocked. Improvements to this check could increase the performance of the tool. Despite this, we still achieve good results.

A second limitation of this approach is the way that localised loops can form between networks of filtered content. This is a key issue with any webcrawling system and often requires human interaction to break the loops - large search engines offer the ability for webmasters to provide links to new sites to improve reach. The looping behaviour we encounter can reduce the effectiveness of the system since the crawler does not have a means to connect other networks of filtered sites. Currently, this can only be altered by manipulating the seed URLs, but is not a fundamental issue with the approach. For purposes of testing and evaluation, limits were not imposed on the traversal between different domains and webpages, but a future implementation could handle looping behaviours in a similar way that search engines avoid spider traps \cite{spidertraps}.

\section{Conclusions}
This work has presented a new approach for building domain filter lists. We demonstrate the method is effective and capable at discovering censored web content in multiple different countries. Given the recursive nature of this method, we envisage that it will be a useful tool for organisations who maintain lists of blocked URLs. Furthermore, the system does not require large amounts of infrastructure to operate or special access to third-party systems or APIs. The use of DNS as a means of checking for filtering has scope to be improved, however, it allows us to test the effectiveness of these kinds of techniques, without incurring ethical issues in regards to the safety of individuals.

Through experimentation on four censorship regimes, we have discovered a large number of filtered domains that have not been previously published. This information will be of significant benefit to future studies concerning research within this field and for organisations that build circumvention tools. We aim to release this data as soon as possible.

Our analysis of the collected data shows the relationship between backlinks of filtered webpages and hyperlinks to other filtered pages. This shows there is indeed a networking effect between different pieces of filtered content and provides a basis for future investigation. Furthermore, our analysis of the types and locations of content being blocked gives insight into the current state of Internet censorship within these regimes.

\section{Future Work}
The approach described in this paper lends itself to refinement and extension. Firstly, the method of checking the filter status of URLs could be improved so it takes into account more factors than only DNS, although care will need to be taken to limit potential harm to people inside censored regions of the world. This could improve the accuracy of the system and potentially increase the scope within which it can operate. 

Secondly, the technique could be integrated with others to form a hybrid system. This may improve performance and reduce the reliance on individual networks of filtered URLs. For example, the search engine based method used by \cite{Darer2017a} would integrate well with this approach. A combined system of this type could improve both the breadth and depth of discovery for filtered URLs by traversing hyperlinks as well as making web searches. Furthermore, this may reduce the closed looping behaviour of solely web crawling.

Thirdly, the data collected by traversing between filtered URLs has potential for further analysis and experimentation. We have touched upon the connectivity between filtered URLs, but there is opportunity for deeper investigation into this concept.

\bibliographystyle{ACM-Reference-Format}
\bibliography{bibli} 

\end{document}